\documentclass[%
 aip,
 amsmath,amssymb,
reprint,%
]{revtex4-2}

\usepackage{graphicx}
\usepackage{dcolumn}
\usepackage{bm}
\usepackage[mathlines]{lineno}
\newcommand*\patchAmsMathEnvironmentForLineno[1]{%
  \expandafter\let\csname old#1\expandafter\endcsname\csname #1\endcsname
  \expandafter\let\csname oldend#1\expandafter\endcsname\csname end#1\endcsname
  \renewenvironment{#1}%
     {\linenomath\csname old#1\endcsname}%
     {\csname oldend#1\endcsname\endlinenomath}}%
\newcommand*\patchBothAmsMathEnvironmentsForLineno[1]{%
  \patchAmsMathEnvironmentForLineno{#1}%
  \patchAmsMathEnvironmentForLineno{#1*}}%
\AtBeginDocument{%
\patchBothAmsMathEnvironmentsForLineno{equation}%
\patchBothAmsMathEnvironmentsForLineno{align}%
\patchBothAmsMathEnvironmentsForLineno{flalign}%
\patchBothAmsMathEnvironmentsForLineno{alignat}%
\patchBothAmsMathEnvironmentsForLineno{gather}%
\patchBothAmsMathEnvironmentsForLineno{multline}%
}

\usepackage{mathptmx}
\usepackage{here}
\usepackage{color}

\begin{document}

\title{Critical comparison of collisionless fluid models: Nonlinear simulations of parallel firehose instability}

\author{Taiki Jikei}
 \email{jikei@eps.s.u-tokyo.ac.jp}
\author{Takanobu Amano}
 \affiliation{Department of Earth and Planetary Science, The University of Tokyo, 7-3-1 Hongo, Bunkyo-ku, Tokyo 113-0033, Japan}

\date{\today}

\begin{abstract}
Two different fluid models for collisionless plasmas are compared. One is based on the classical Chew-Goldberger-Low (CGL) model that includes a finite Larmor radius (FLR) correction and the Landau closure for the longitudinal mode. Another one takes into account the effect of cyclotron resonance in addition to Landau resonance, which is referred to as the cyclotron resonance closure (CRC) model \citep{Jikei2021}. While the linear property of the parallel firehose instability is better described by the CGL model, the electromagnetic ion cyclotron instability driven unstable by the cyclotron resonance is reproduced only by the CRC model. Nonlinear simulation results for the parallel firehose instability performed with the two models are also discussed. Although the linear and quasilinear isotropization phases are consistent with theory in both models, long-term behaviors may be substantially different. The final state obtained by the CRC model may be reasonably understood in terms of the marginal stability condition. In contrast, the lack of cyclotron damping in the CGL model makes it rather difficult to predict the long-term behavior with a simple physical argument. This suggests that incorporating the collisionless damping both for longitudinal and transverse modes is crucial for a nonlinear fluid simulation model of collisionless plasmas.
\end{abstract}

\maketitle

\section{Introduction}
Compared to fully kinetic models that solve the Vlasov equation directly on up to six-dimensional phase space for at least one of the species, fluid modeling of collisionless plasmas requires much less computational resources since only finite numbers of fluid moment equations are involved, such as the continuity equation and the equation of motion (EoM).
In addition, it is much easier to discuss the macroscopic dynamics only in terms of moment quantities without considering the complicated shape of the distribution function.
Magnetohydrodynamics (MHD), which is the most commonly used fluid model at present, assumes a scalar pressure and typically adopts various simplifications of the equation of state (EoS). It thus ignores the collisionless wave-particle interaction completely even though it is well-known that wave-particle interaction may have non-negligible impacts even at macroscopic scales.
Clearly, a fluid model that properly takes into account wave-particle interaction is needed for macroscopic modeling of the collisionless plasma dynamics.

The Chew-Goldberger-Low (CGL) model \citep{Chew1956} is one of the examples of collisionless fluid models that takes into account pressure anisotropy. It assumes that the time scale of the phenomena is much longer than the inverse ion cyclotron frequency and may be considered as a natural extension of MHD. Various extensions of the original CGL model have been proposed, and the abilities to reproduce the linear kinetic theory have been investigated.
Two major improvements of particular note are the finite Larmor radius (FLR) correction and the Landau closure to the longitudinal mode \citep{Hunana2019part1,Hunana2019part2}.

We have recently developed a different kind of collisionless fluid model. It solves the equation of the full pressure tensor without explicitly introducing the low-frequency approximation \citep{Jikei2021} and is called the cyclotron resonance closure (CRC) model in this paper.
Following the original approach of the Landau closure for electrostatic modes \citep{Hammett1990,Hammett1992}, we approximate the heat flux components associated with electromagnetic waves by a linear combination of lower-order moments in Fourier space. The resulting model approximately reproduces the linear cyclotron resonance effect, both the cyclotron damping and the instability of electromagnetic ion cyclotron (EMIC) waves in the presence of a sufficiently large perpendicular temperature anisotropy. A nonlinear simulation of the EMIC instability with the CRC model shows a reasonable behavior both in linear and fully nonlinear phases.

The purpose of this paper is to present a detailed comparison between the two models, both in the linear and nonlinear properties. We show that the CGL model better describes the parallel firehose instability than the CRC model. On the other hand, the EMIC instability is not reproduced by the CGL model as it lacks the essential cyclotron resonance effect. Although the small degrees of freedom in the moment expansion of the CRC model limits the accuracy in describing the linear wave property, it predicts a qualitatively correct behavior for a wide range of parameters. For the nonlinear problem, we consider the parallel firehose instability that can be reproduced by both the CGL and CRC models. We find that the linear and quasilinear evolutions are consistent with the theoretical predictions described in Sec.~\ref{sec2}. On the other hand, the long-term fully nonlinear states may be controlled by the dissipation incorporated in the model. We find that the CRC model gives a more physically reasonable final state that is described by the marginal stability condition. We attribute it to the presence of the collisionless damping of both the transverse and longitudinal modes. In contrast, the absence of the linear damping of the transverse mode in the CGL model may lead to unphysical behaviors in the fully nonlinear state. Therefore, we conclude that the CRC model is better suited for a nonlinear simulation model even though the CGL model may provide an excellent approximation to some of the linear kinetic instability.

The remainder of this paper is organized as follows. We introduce the two collisionless fluid models in Sec. \ref{sec2}. We also discuss quasilinear isotropization of pressure anisotropy and Landau damping of sound waves, which are important to understand the nonlinear simulation results. In Sec. \ref{sec3}, we consider proton temperature anisotropy instability, in particular, the parallel firehose instability, to compare the two models. Both the linear theory and the nonlinear simulation results are discussed in detail. Finally, discussion and conclusions will be given in Sec. \ref{sec4}.

\section{Fluid description of collisionless plasmas} \label{sec2}

\subsection{Moment hierarchy}
The dynamics of collisionless plasmas is governed by the Vlasov equation
\begin{equation} \label{Vlasov}
\frac{\partial f_s}{\partial t}+\bm{v}\cdot\nabla f_s+\frac{e_s}{m_s}\left(\bm{E}+\bm{v}\times\bm{B}\right)
\cdot\frac{\partial f_s}{\partial \bm{v}}=0,
\end{equation}
which describes the time evolution of the distribution function $f_s(\bm{x},\bm{v};t)$ for the species $s$ as a function of position $\bm{x}$, velocity $\bm{v}$ and time $t$ ($s=p$ for protons and $s=e$ for electrons) under the action of the electric and magnetic fields $\bm{E}$ and $\bm{B}$. Here, $e_s$ and $m_s$ denote the particle charge and mass, respectively. Starting from Eq. (\ref{Vlasov}), we can derive the moment equations using the number density $n_s=\int f_s d^3v$, fluid velocity $\bm{u}_s=\int \bm{v} f_s d^3v/n_s$, pressure tensor $\bm{p}_s= m_s\int (\bm{v}-\bm{u}_s)(\bm{v}-\bm{u}_s)f_s d^3v$ and heat flux tensor $\bm{q}_s= m_s\int (\bm{v}-\bm{u}_s)(\bm{v}-\bm{u}_s)(\bm{v}-\bm{u}_s)f_s d^3v$, etc. The equations up to the second-order moment are written as follows
\begin{align}
& \frac{\partial n_s}{\partial t}+\nabla\cdot(n_s\bm{u}_s)=0, \label{continuity} \\
& \frac{\partial \bm{u}_s}{\partial t}+\bm{u}_s\cdot\nabla\bm{u}_s+\frac{1}{m_sn_s}\nabla\cdot\bm{p}_s-\frac{e_s}{m_s}(\bm{E}+\bm{u}_s\times\bm{B})=\bm{0}, \label{EOM} \\
& \frac{\partial \bm{p}_s}{\partial t}+\nabla\cdot(\bm{p}_s\bm{u}_s+\bm{q}_s)+\left(\bm{p}_s\cdot\nabla\bm{u}_s+\frac{e_s}{m_s}\bm{B}\times\bm{p}_s\right)^{\mathcal{S}}=\bm{0}. \label{pressureeq}
\end{align}
We define the divergence of a tensor with respect to the last index, i.e., $\nabla\cdot\bm{A}=\partial_nA_{ijk\ldots n}$. The superscript $\mathcal{S}$ denotes a symmetrization operator $\bm{A}^{\mathcal{S}}=\bm{A}+\bm{A}^\top$.

The evolution of the electromagnetic field is governed by Maxwell's equations
\begin{align}
& \nabla\cdot\bm{E}=\frac{\rho_e}{\epsilon_0}, \label{MW1} \\
& \nabla\cdot\bm{B}=0, \label{MW2} \\
& \nabla\times\bm{E}=-\frac{\partial \bm{B}}{\partial t}, \label{MW3} \\
& \nabla\times\bm{B}=\mu_0\left(\bm{j}+\epsilon_0\frac{\partial \bm{E}}{\partial t}\right), \label{MW4}
\end{align}
where $\epsilon_0$ and $\mu_0$ are the permittivity and permeability of vacuum.
For electron-proton plasmas with $e_p=+e,\, e_e=-e$, the charge density $\rho_e$ and the current density $\bm{j}$ are calculated with the fluid quantities
\begin{align}
& \rho_e=e(n_p-n_e), \label{charge} \\
& \bm{j}=e(n_p\bm{u}_p-n_e\bm{u}_e), \label{current}
\end{align}
where $e$ is the elementary charge.

Note that the set of equations are not yet closed unless the heat flux $\bm{q}_s$ is given. Therefore, we have to introduce a model to determine the heat flux either explicitly or implicitly. Although we focus on modeling the heat flux in this paper, the need for a closure model is generic. It is well known that a moment quantity of order $n+1$ always appears in the time evolution equation for the $n$-th order moment. To obtain a closed set of equations, a closure assumption has to be made to determine the highest order moment.

Although the closure model may, in general, be introduced independently for each species, in this paper, we consider only the dynamics of protons, assuming that the electrons are a charge-neutralizing massless cold fluid. This choice is motivated because many CGL-based collisionless fluid models are essentially extensions of Hall-MHD. In other words, we consider only low-frequency phenomena relative to the electron plasma and cyclotron frequencies. We do not, however, rule out any phenomena around and even beyond the proton cyclotron frequency at this point.

Hereafter, the subscript $p$ will be omitted because the fluid quantities always refer to those of protons. We ignore the displacement current in Eq. (\ref{MW4}), and the current density is given by
\begin{equation} \label{nodc}
\bm{j}=\frac{\nabla\times\bm{B}}{\mu_0}.
\end{equation}
Instead of solving Eq. (\ref{MW1}), the electric field is given by Ohm's law
\begin{equation} \label{Ohm}
\bm{E}=-\bm{u}\times\bm{B}+\frac{\bm{j}\times\bm{B}}{ne}.
\end{equation}
Note that we have ignored the divergence of the electron pressure tensor because electrons are cold. Substituting these into Eq. (\ref{EOM}) and Eq. (\ref{MW3}), we obtain the single-fluid EoM
\begin{equation} \label{mfEOM}
\frac{\partial \bm{u}}{\partial t}+\bm{u}\cdot\nabla\bm{u}+\frac{1}{mn}\nabla\cdot\bm{p}-\frac{1}{\mu_0mn}(\nabla\times\bm{B})\times\bm{B}=\bm{0},
\end{equation}
and the magnetic induction equation
\begin{equation} \label{induction}
\frac{\partial \bm{B}}{\partial t}=\nabla\times(\bm{u}\times\bm{B})-\frac{1}{\mu_0e}\nabla\times\left[\frac{(\nabla\times\bm{B})\times\bm{B}}{n}\right].
\end{equation}
It is easy to confirm that the total energy of the system given by
\begin{equation} \label{energy}
\frac{1}{2}\mathrm{Tr} \bm{p} + \frac{1}{2}mnu^2+\frac{B^2}{2\mu_0}
\end{equation}
is a conserved quantity. Note that the model is equivalent to the standard Hall-MHD if the adiabatic equation of state is used for a scalar pressure and the heat flux is simply neglected.

\subsection{The CGL model} \label{CGLeqs}
The original CGL model \citep{Chew1956} in the absence of heat flux and FLR corrections is the simplest collisionless fluid model under the low-frequency approximation. With an appropriate FLR correction, it gives an adequate approximation to both parallel and oblique firehose instabilities. Furthermore, by incorporating a non-local Landau-type closure to the heat flux, it may describe the correct mirror instability threshold as predicted by fully kinetic theory. More details on the CGL models can be found in recent comprehensive reviews \citep{Hunana2019part1,Hunana2019part2}. For nonlinear simulations shown in this paper, we employ a simple linear FLR correction to the pressure tensor, which is called FLR1 by \citet{Hunana2019part1}. Although more sophisticated FLR correction models have been proposed including fully nonlinear ones \citep{Ramos2005,Goswami2005,Passot2004,Passot2007,Sulem2015}, we think that this FLR model is sufficient for our purpose of comparison with the CRC model that will be introduced in the next subsection.

In the following, we consider a homogeneous plasma and define the ambient magnetic field to be along $z$ direction $\bm{B}_0=B_0\bm{\hat{e}}_z$.   Introducing a small parameter $\epsilon \sim \omega/\Omega \sim k r_g$ (where $\Omega = e B_0/m$ is the cyclotron frequency, $k$ is the wavenumber, and $r_g$ is the Larmor radius), we expand Eq.~(\ref{pressureeq}) in powers of $\epsilon$. At the lowest order $O(\epsilon^{0})$, we have
\begin{equation} \label{lfpeq}
\Omega\frac{|\bm{B}|}{B_0}(\bm{\hat{b}}\times\bm{p})^\mathcal{S}=0,
\end{equation}
which describes the fast gyromotion of the particles. Note that the magnetic field unit vector is denoted by $\bm{\hat{b}}=\bm{B}/|\bm{B}|$. In general, the gyrotropic pressure tensor $\bm{P}^g$ defined by
\begin{equation} \label{pgyrotropic}
    \bm{p} = \bm{P}^\mathrm{g} = P_{\parallel}\bm{\hat{b}}\bm{\hat{b}}+P_{\perp}(\bm{I}-\bm{\hat{b}}\bm{\hat{b}}).
\end{equation}
satisfies Eq.~(\ref{lfpeq}). The parallel and perpendicular pressures are given by
\begin{align}
& P_{\parallel}=\bm{p}:\bm{\hat{b}}\bm{\hat{b}}, \label{ppara} \\
& P_{\perp}=\bm{p}:(\bm{I}-\bm{\hat{b}}\bm{\hat{b}})/2, \label{pperp}
\end{align}
with $\bm{I}$ being the unit tensor and $:$ denoting the double contraction operator $\bm{A}:\bm{B}=A_{ij}B_{ij}$.

We now define a non-gyrotropic FLR correction to the pressure tensor $\bm{\Pi}$ by
\begin{equation} \label{decomposition}
\bm{p}= \bm{P}^\mathrm{g} + \bm{\Pi}.
\end{equation}
It is immediately found that the FLR correction must be of $O(\epsilon)$ to be consistent with Eq.~(\ref{lfpeq}). In the following, we ignore non-gyrotropic heat flux and consider terms of $O(\epsilon)$ in Eq.~(\ref{pressureeq}). By applying the double contraction with $\bm{\hat{b}}\bm{\hat{b}}$, we have the parallel pressure equation
\begin{equation} \label{pparaeq}
\frac{\partial P_{\parallel}}{\partial t}+\nabla\cdot(P_{\parallel}\bm{u}+Q_{\parallel}\bm{\hat{b}})+2P_{\parallel}\bm{\hat{b}}\cdot\nabla\bm{u}\cdot\bm{\hat{b}}-2Q_{\perp}\nabla\cdot\bm{\hat{b}}=0.
\end{equation}
Similarly, for the perpendicular pressure, we have
\begin{equation} \label{pperpeq}
\frac{\partial P_{\perp}}{\partial t}+\nabla\cdot(P_{\perp}\bm{u}+Q_{\perp}\bm{\hat{b}})+P_{\perp}\nabla\cdot\bm{u}-P_{\perp}\bm{\hat{b}}\cdot\nabla\bm{u}\cdot\bm{\hat{b}}+2Q_{\perp}\nabla\cdot\bm{\hat{b}}=0.
\end{equation}
Note that the FLR correction terms in the above equations have been dropped as they are of higher order in $\epsilon$. The gyrotropic heat flux components $Q_{\parallel}, \, Q_{\perp}$ will be specified later. We will denote the pressure and heat flux components after the decomposition with respect to the local magnetic field direction by capital letters, e.g., $P_{\parallel}, \, P_{\perp}, \, Q_{\parallel}, \, Q_{\perp}$ (with subscripts $\parallel, \, \perp$). On the other hand, the pressure and heat flux tensor components in the original Cartesian coordinate will be expressed by lower cases, e.g., $p_{xz}, \, q_{xzz}.$

Remaining components of $O(\epsilon)$ in Eq.~(\ref{pressureeq}) give the following simple linear FLR correction that will be used in this paper:
\begin{equation} \label{FLR1}
\begin{cases}
\Pi_{xx}=-\Pi_{yy}=-\frac{P_{\perp0}}{2\Omega}\left(\frac{\partial u_y}{\partial x}+\frac{\partial u_x}{\partial y}\right), \\
\Pi_{xy}=\frac{P_{\perp0}}{2\Omega}\left(\frac{\partial u_x}{\partial x}-\frac{\partial u_y}{\partial y}\right), \\
\Pi_{xz}=-\frac{1}{\Omega}\left[(2P_{\parallel0}-P_{\perp0})\frac{\partial u_y}{\partial z}+P_{\perp0}\frac{\partial u_z}{\partial y}\right], \\
\Pi_{yz}=\frac{1}{\Omega}\left[(2P_{\parallel0}-P_{\perp0})\frac{\partial u_x}{\partial z}+P_{\perp0}\frac{\partial u_z}{\partial x}\right], \\
\Pi_{zz}=0.
\end{cases}
\end{equation}
Note that we evaluate $P_{\parallel0}$, $P_{\perp0}$, and $\Omega$ in the correction by averaging these quantities over the simulation box at each time step.

For the gyrotropic heat flux components, we use the closure proposed by \citep{Snyder1997}
\begin{align}
& Q_{\parallel}=-n_0\sqrt{\frac{8}{\pi}}V_{\mathrm{th}}\mathcal{H}[T_{\parallel}], \label{snyderpara} \\
& Q_{\perp}=-n_0\sqrt{\frac{2}{\pi}}V_{\mathrm{th}}\mathcal{H}[T_{\perp}]+n_0\sqrt{\frac{2}{\pi}}V_{\mathrm{th}}T_{\perp 0}\left(1-\frac{T_{\perp 0}}{T_{\parallel 0}}\right)\mathcal{H}\left[\frac{|\bm{B}|}{B_0}\right]. \label{snyderperp}
\end{align}
The parallel, perpendicular temperatures, and the parallel thermal velocity are respectively denoted by $T_{\parallel}=P_{\parallel}/n, \, T_{\perp}=P_{\perp}/n$ and $V_{\mathrm{th}}=\sqrt{T_{\parallel 0}/m}.$
We have defined the (negative) Hilbert transform $\mathcal{H}$, which, in Fourier space, is written as $\mathcal{H}=\frac{ik_z}{|k_z|}$ with the parallel wavenumber $k_z$.
Hilbert transform needs integration along the ambient field, or equivalently Fourier transform, which makes this closure non-local in the sense that the heat flux cannot be calculated by local quantities and their derivatives of finite orders alone.

Although it is a model for collisionless plasmas, later we will find it useful to introduce an effective resistivity. 
We thus modify Eq.~(\ref{induction}) to
\begin{equation} \label{inductionres}
\begin{split}
\frac{\partial \bm{B}}{\partial t}=&\nabla\times(\bm{u}\times\bm{B})-\frac{1}{\mu_0e}\nabla\times\left[\frac{(\nabla\times\bm{B})\times\bm{B}}{n}\right]\\
&-\frac{B_0}{\mu_0n_0e}\nabla\times(\eta\nabla\times\bm{B}),
\end{split}
\end{equation}
where $\eta$ is the dimensionless resistivity. It is the reciprocal of the magnetic Reynolds number defined with typical velocity and length scales set to the Alfv\`{e}n speed and proton inertial length, respectively. For the conservation of energy, we also modify Eqs.~(\ref{pparaeq}) and (\ref{pperpeq}) as follows
\begin{equation}
\begin{split}
\frac{\partial P_{\parallel}}{\partial t}&+\nabla\cdot(P_{\parallel}\bm{u}+Q_{\parallel}\bm{\hat{b}})+2P_{\parallel}\bm{\hat{b}}\cdot\nabla\bm{u}\cdot\bm{\hat{b}}\\
&-2Q_{\perp}\nabla\cdot\bm{\hat{b}}=\frac{2B_0}{3\mu_0n_0e}\eta\left|\nabla\times\bm{B}\right|^2, \label{pparaeqres}
\end{split}
\end{equation}
\begin{equation}
\begin{split}
\frac{\partial P_{\perp}}{\partial t}&+\nabla\cdot(P_{\perp}\bm{u}+Q_{\perp}\bm{\hat{b}})+P_{\perp}\nabla\cdot\bm{u}-P_{\perp}\bm{\hat{b}}\cdot\nabla\bm{u}\cdot\bm{\hat{b}}\\
&+2Q_{\perp}\nabla\cdot\bm{\hat{b}}=\frac{2B_0}{3\mu_0n_0e}\eta\left|\nabla\times\bm{B}\right|^2, \label{pperpeqres}
\end{split}
\end{equation}
where we have assumed that Ohmic heating gives the equal amount of energy to the particles for each degree of freedom. Note that the effective resistivity is not able to determine the energy partition in a self-consistent manner. In contrast, we will see in the next subsection that the CRC model consistently takes into account the dissipation of electromagnetic wave energy via the cyclotron damping.
In this paper, the CGL model refers to the model with the Hall term, the FLR correction of the form of Eq. (\ref{FLR1}), and the Landau closure Eqs. (\ref{snyderpara},\ref{snyderperp}), unless otherwise noted.
It is to be noted that compared to the original CGL model in which the heat flux is ignored, the finite $Q_{\parallel}, Q_{\perp}$ break the conservation of the first and second adiabatic invariants.

\subsection{The CRC model} \label{CRCmodel}
We have recently proposed a kinetic fluid model that takes into account the effect of cyclotron resonance for electromagnetic waves propagating along the ambient magnetic field, which we call the CRC model \citep{Jikei2021}. It correctly reproduces both linear growth and damping of transverse fluctuations through the cyclotron resonance. Furthermore, we have shown that nonlinear simulations for the EMIC instability driven by a proton temperature anisotropy adequately describe the reduction of initial anisotropy and the saturation of wave growth.

Let us briefly describe the CRC model. We consider a strictly one-dimensional system along the ambient magnetic field for simplicity. Assuming the symmetry of the pressure tensor in the plane perpendicular to the ambient field
\begin{align}
p_{xx}=p_{yy}, \label{pxxyy} \\
p_{xy}=0, \label{pxy}
\end{align}
we compute the time evolution of the full pressure tensor with Eq. (\ref{pressureeq}).

The four independent components of the pressure equation read
\begin{equation}
\begin{split}
\frac{\partial p_{zz}}{\partial t}&+\frac{\partial}{\partial z}\left(p_{zz}u_z+q_{zzz}\right)\\
&+2\left[p_{zz}\frac{\partial u_z}{\partial z}+\frac{e}{m}(B_xp_{yz}-B_yp_{xz})\right]=0, \label{nlppara}
\end{split}
\end{equation}
\begin{equation}
\begin{split}
\frac{\partial p_{xx}}{\partial t}&+\frac{\partial}{\partial z}\left(p_{xx}u_z + q_{xxz} + q_{yyz} \right)+p_{xz}\frac{\partial u_x}{\partial z}+p_{yz}\frac{\partial u_y}{\partial z}\\
&-\frac{e}{m}(B_xp_{yz}-B_yp_{xz})=0, \label{nlpperp}
\end{split}
\end{equation}
\begin{equation}
\begin{split}
\frac{\partial p_{xz}}{\partial t}&+\frac{\partial}{\partial z}(p_{xz}u_z+q_{xzz})+p_{xz}\frac{\partial u_z}{\partial z}+p_{zz}\frac{\partial u_x}{\partial z}\\
&+\frac{e}{m}[B_y(p_{zz}-p_{xx})-B_0p_{yz}]=0, \label{pxz}
\end{split}
\end{equation}
\begin{equation}
\begin{split}
\frac{\partial p_{yz}}{\partial t}&+\frac{\partial}{\partial z}(p_{yz}u_z+q_{yzz})+p_{yz}\frac{\partial u_z}{\partial z}+p_{zz}\frac{\partial u_y}{\partial z}\\
&+\frac{e}{m}[B_x(p_{xx}-p_{zz})+B_0p_{xz}]=0. \label{pyz}
\end{split}
\end{equation}
Note that Eq.~(\ref{nlpperp}) is obtained by taking the average of equations for $p_{xx}$ and $p_{yy}$ and using the assumed symmetry. To obtain a closed set of equations, we have to adopt a model for the heat flux components $q_{xxz}, q_{yyz}, q_{xzz}, q_{yzz}, q_{zzz}$.
For the diagonal components, we use
\begin{align}
q_{zzz}=-n_0\sqrt{\frac{8}{\pi}}v_{\mathrm{th}}\mathcal{H}[T_{zz}], \label{qzzz} \\
q_{xxz}+q_{yyz}=0. \label{qxxyyz}
\end{align}
Note that $q_{zzz}$ is identical to $Q_{\parallel}$ except for the definition of parallel temperature $T_{zz}=p_{zz}/n$ and thermal velocity $v_{\mathrm{th}}=\sqrt{T_{zz0}/m}$. On the other hand, the assumption of $q_{xxz}+q_{yyz} = 0$ is different from the assumption of $Q_{\perp}$. 
The reason for this is that the CRC model considers a linearized closure with respect to the ambient magnetic field, which completely decouples the parallel and perpendicular dynamics. In this case, the perpendicular pressure must be independent of Landau damping of longitudinal fluctuations.
To calculate the off-diagonal components, we decompose the transverse quantities with $\hat{q}=q_{xzz}+iq_{yzz},\,\hat{p}=p_{xz}+ip_{yz},\,\hat{u}=u_x+iu_y$, and use the closure model \citep{Jikei2021}
\begin{equation} \label{cyclotronclosure}
\hat{q}=-\frac{\sqrt{2\pi}}{\pi-2}v_{\mathrm{th}}\mathcal{H}[\hat{p}]+\frac{4-\pi}{\pi-2}p_{zz0}\hat{u}.
\end{equation}
Note that $q_{xzz}$ and $q_{yzz}$ are recovered as the real and imaginary parts of $\hat{q}$, respectively.

It is important to point out that we did not introduce any assumptions on the time scale. Therefore, the model may well describe the phenomena even beyond the ion cyclotron frequency as long as the closure itself is adequate. This is in clear contrast with the CGL model, in which the model, in a strict sense, should be appropriate only to first order in $\epsilon \sim \omega/\Omega \sim k r_g$. Another advantage is that it explicitly contains the energy exchange with transverse waves by the term $p_{xz}\frac{\partial u_x}{\partial z}+p_{yz}\frac{\partial u_y}{\partial z}$ in Eq.~(\ref{nlpperp}). For instance, if the cyclotron damping dissipates the wave energy, the energy will be converted to the perpendicular pressure. As will be shown in the next subsection, there may be further energy exchange between parallel and perpendicular directions, which is driven by the term proportional to $(B_xp_{yz}-B_yp_{xz})$ in Eqs.~(\ref{nlppara}-\ref{nlpperp}).

\subsection{Quasilinear isotropization} \label{quasilinear}
Although a fluid model cannot describe the full nonlinearity of the Vlasov system which possesses an infinite number of conserved quantities, we will show that quasilinear temperature isotropization (energy exchange between parallel and perpendicular directions) by transverse electromagnetic waves is included in a model.

We consider parallel propagating transverse waves and rewrite Eq. (\ref{pparaeq}) with $\partial/\partial z$
\begin{equation} \label{ppara2}
\begin{split}
\frac{\partial P_{\parallel}}{\partial t}+\frac{\partial}{\partial z}(P_{\parallel}u_z+q_{\parallel}\hat{b}_z)&+2P_{\parallel}\hat{b}_z\left(\hat{b}_x\frac{\partial u_x}{\partial z}+\hat{b}_y\frac{\partial u_y}{\partial z}+\hat{b}_z\frac{\partial u_z}{\partial z}\right)\\
&-2Q_{\parallel}\frac{\partial \hat{b}_z}{\partial z}=0.
\end{split}
\end{equation}
We are interested in the time evolution of spatially averaged quantities, for which the divergence term is not relevant.
Since we only consider transverse waves, we assume $\left|\frac{\partial u_z}{\partial z}\right|\ll\left|\frac{\partial u_{x,y}}{\partial z}\right|$.
The linear approximation of $\bm{\hat{b}}$
\begin{equation} \label{bhat}
\bm{\hat{b}}\approx
\bm{\hat{b}}_0 +
\begin{pmatrix}
    \frac{B_x}{B_0} \\
    \frac{B_y}{B_0} \\
    0
\end{pmatrix}
=    
\begin{pmatrix}
\frac{B_x}{B_0} \\
\frac{B_y}{B_0} \\
1
\end{pmatrix}
\end{equation}
indicates that $Q_{\parallel}\frac{\partial \hat{b}_z}{\partial z}$ is negligible at the second-order level.
Therefore, the time evolution of the average parallel pressure can be written in the convolution form
\begin{equation} \label{pparaconv}
\frac{\partial P_{\parallel 0}}{\partial t}\approx-\frac{2iP_{\parallel 0}}{B_0}\int k[B_x(-k)u_x(k)+B_y(-k)u_y(k)]dk.
\end{equation}
If we ignore the Hall term, the linear phase relation of transverse velocity and magnetic field reads
\begin{align}
\tilde{u}_x=-\frac{\omega}{k}\frac{\tilde{B}_x}{B_0}, \\
\tilde{u}_y=-\frac{\omega}{k}\frac{\tilde{B}_y}{B_0}.
\end{align}
Then we can rewrite Eq. (\ref{pparaconv}) as
\begin{equation} \label{pparaconv2}
\frac{\partial P_{\parallel 0}}{\partial t}\approx2iP_{\parallel 0}\int \omega(k)\frac{|\tilde{B}_x(k)|^2+|\tilde{B}_y(k)|^2}{B^2_0}dk.
\end{equation}
Given the linear growth rate and the amplitude of the magnetic fluctuations, we can evaluate the isotropization during the quasilinear phase.
Similarly, we can obtain the time evolution of perpendicular pressure from the term $P_{\perp}\bm{\hat{b}}\cdot\nabla\bm{u}\cdot\bm{\hat{b}}$ in Eq. (\ref{pperpeq}). For more general CGL models, the isotropization term may be evaluated by substituting the linear eigenvector in Eq.~(\ref{pparaconv}).

As was shown in \citep{Jikei2021}, the isotropization in the CRC model comes from the term proportional to $(B_xp_{yz}-B_yp_{xz})$. The convolution form of Eq. (\ref{nlppara}) for transverse waves up to second-order reads
\begin{equation} \label{pzzconv}
\frac{\partial p_{zz0}}{\partial t}\approx-\frac{2 e}{m}\int [\tilde{B}_x(-k)\tilde{p}_{yz}(k)-\tilde{B}_y(-k)\tilde{p}_{xz}(k)]dk.
\end{equation}
We demonstrated that the isotropization observed in the quasilinear phase of a nonlinear simulation was consistent with the theoretical estimate.

While it might appear that the isotropization terms in the CGL and CRC models arise from different physical origins, we can show that they are indeed consistent with each other.
To see this, recall that parallel pressure equation with CGL model was obtained by applying $:\bm{\hat{b}}\bm{\hat{b}}$ to Eq. (\ref{pressureeq}).
By applying $:\bm{\hat{b}}\bm{\hat{b}}$ to the pressure tensor equation Eq.~(\ref{pressureeq}) and dropping higher-order terms and the $\nabla\cdot u$ term, we obtain
\begin{equation} \label{pparapw}
\begin{split}
&\left[\frac{\partial p_{zz}}{\partial t}+\frac{2B_x}{B_0}\frac{\partial p_{xz}}{\partial t}+\frac{2B_y}{B_0}\frac{\partial p_{yz}}{\partial t}\right]\\
&+\frac{\partial}{\partial z}(p_{zz}u_z+q_{zzz})+2p_{zz}\left(\frac{B_x}{B_0}\frac{\partial u_x}{\partial z}+\frac{B_y}{B_0}\frac{\partial u_y}{\partial z}\right)=0.
\end{split}
\end{equation}
We find that the first term corresponds to the CGL parallel pressure time derivative since
\begin{equation}
\frac{\partial P_{\parallel}}{\partial t}=\frac{\partial}{\partial t}(\bm{\hat{b}}\bm{\hat{b}}):\bm{p}+\bm{\hat{b}}\bm{\hat{b}}:\frac{\partial \bm{p}}{\partial t}
\approx \bm{\hat{b}}\bm{\hat{b}}:\frac{\partial \bm{p}}{\partial t},
\end{equation}
where $\frac{\partial}{\partial t}(\bm{\hat{b}}\bm{\hat{b}})$ is negligible in the CGL approximation \citep{Hunana2019part1}. On the other hand, it is straightforward to show that the following relation is satisfied from Eqs.~(\ref{pxz}--\ref{pyz})
\begin{align} \label{relation-of-isotropization}
    \frac{B_x}{B_0} \frac{\partial p_{xz}}{\partial t} +
    \frac{B_y}{B_0} \frac{\partial p_{yz}}{\partial t} +
    p_{zz} \left(
        \frac{B_x}{B_0} \frac{\partial u_x}{\partial z} +
        \frac{B_y}{B_0} \frac{\partial u_y}{\partial z}
    \right) =
    \frac{e}{m} \left( B_x p_{yz} - B_y p_{xz} \right),
\end{align}
if the spatial average is taken and only transverse fluctuations are considered. We thus find that the quasilinear evolution of pressure given by Eq.~(\ref{pparaconv}) for CGL and Eq.~(\ref{pzzconv})  for CRC are fully consistent with each other. We interpret this result as follows. Noting that $(\bm{B}\times\bm{p})^{\mathcal{S}}:\bm{\hat{b}}\bm{\hat{b}} = 0$ is satisfied exactly by definition, the isotropization associated with the term $(\bm{B}\times\bm{p})^{\mathcal{S}}$ in the original coordinate disappears after the decomposition to the gyrotropic pressures $P_{\parallel}$ and $P_{\perp}$. Instead, the $\bm{B}\cdot\nabla\bm{u}$ term that is driven finite by the off-diagonal pressure components via Eq.~(\ref{relation-of-isotropization}) plays the role of isotropization.

These results indicate that both the CGL and CRC models are capable of reproducing the quasilinear isotropization during the early nonlinear stage where the phase relation between the magnetic field and the fluid quantities remain consistent with the linear eigenvector. The difference between the models in the quasilinear phase should thus result from the difference in the linear wave property but not from the treatment of the pressure tensor.

\subsection{Landau damping of sound waves} \label{LD}
Before moving on to the comparison between the CGL model and the CRC model, let us discuss the effect of Landau damping of longitudinal modes, which is practically identical for both models.
Although the parallel firehose instability, which we discuss in the next section, is purely transverse in the linear phase, substantial longitudinal perturbations will be generated in the nonlinear phase.
The Landau damping thus provides an important dissipation channel in the system.

\begin{figure}[H]
\begin{center}
\includegraphics[width=0.85\linewidth]{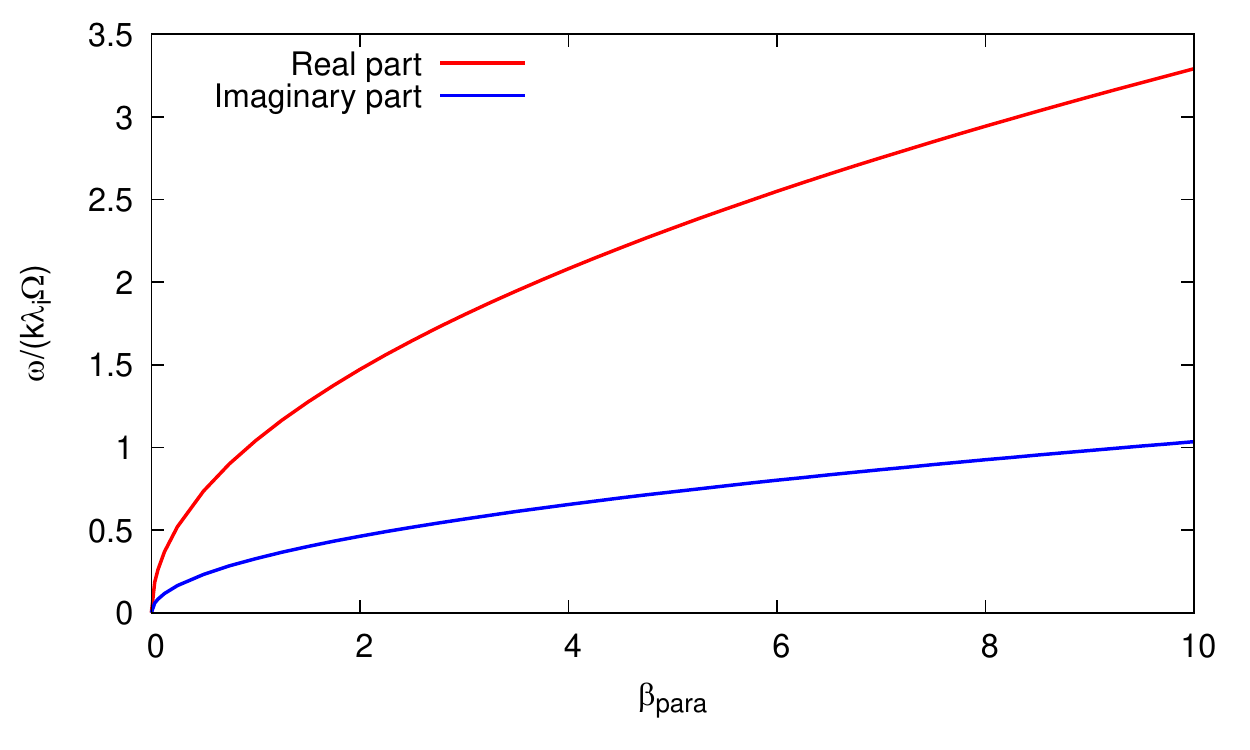}%
\caption{\label{sound} Normalized phase velocity of sound waves. The real and imaginary parts are shown in red and blue respectively. Note that frequency is plotted in red and the damping rate $(-\mathrm{Im}[\omega]/(k\lambda_i\Omega))$ is shown for the imaginary part.}
\end{center}
\end{figure}

The linearized set of equations for one-dimensional electrostatic system reads
\begin{align}
-i\omega\tilde{n}+ikn_0\tilde{u}_z=0,\\
-i\omega\tilde{u}+\frac{ik}{mn_0}\tilde{p}_{zz}=0,\\
-i\omega\tilde{p}_{zz}+ik(3p_0\tilde{u}_z+\tilde{q}_{zzz}),\\
\tilde{q}_{zzz}=-\frac{ik}{|k|}\sqrt{\frac{8}{\pi}}\sqrt{\frac{p_0}{mn_0}}\left(\tilde{p}-p_0\frac{\tilde{n}}{n_0}\right).
\end{align}
The dispersion relation of sound waves can be obtained as the solution of
\begin{equation}
\left(\frac{\omega}{k\lambda_i\Omega}\right)^3+i\sqrt{\frac{4\beta_{\parallel}}{\pi}}\left(\frac{\omega}{k\lambda_i\Omega}\right)^2-\frac{3\beta_{\parallel}}{2}\left(\frac{\omega}{k\lambda_i\Omega}\right)-i\sqrt{\frac{\beta_{\parallel}^3}{\pi}}=0,
\end{equation}
which is shown in FIG. \ref{sound}. The proton inertial length is denoted by $\lambda_i=\sqrt{\frac{m}{\mu_0ne^2}}$. Note that the normalized phase velocity is shown because the sound wave in this system is non-dispersive.
We see that the Landau damping rate is clearly higher at high-beta plasmas.
However, the lack of any characteristic length scales makes it difficult to estimate the scale length at which the linear damping and the steepening of a finite amplitude wave balance with each other unless the amplitude of a sound wave is known beforehand.

\section{Proton temperature anisotropy instability} \label{sec3}
In this section, we consider proton temperature anisotropy instability that excites transverse waves propagating parallel to the ambient magnetic field.
It is well known that when $a=T_{\perp}/T_{\parallel}$ and $\beta_{\parallel}=2\mu_0P_{\parallel}/B_0^2$ (or an equivalent pair of parameters) satisfy certain conditions, a uniform system of collisionless plasma becomes unstable.
If the perpendicular pressure exceeds the parallel pressure, left-handed polarized EMIC waves become unstable (EMIC instability).
On the other hand, if the parallel pressure dominates, right-handed polarized whistler waves are destabilized by the parallel firehose instability.
Sufficiently above the threshold of $\beta_{\parallel}>2$, the firehose instability is non-resonant and may be reproduced by the CGL model\citep{Rosin2011,Schekochihin2010,Wang2010}.
It should be noted that, however, the cyclotron resonance plays a non-negligible role in firehose instability at large wavenumbers or lower $\beta_{\parallel}$\citep{Hunana2017,Hunana2019part1}.
On the other hand, the cyclotron resonance is indispensable for reproducing the EMIC instability, for which the CGL model is clearly inadequate.
We discuss the firehose instability since it can be reproduced by both the CGL and the CRC models.
Our goal in this section is to investigate whether or not these two models are appropriate for describing the nonlinear evolution of plasma instabilities driven by temperature anisotropy.

\subsection{Linear theory}
The dispersion relation of a transverse wave propagating parallel to the ambient magnetic field with fully kinetic bi-Maxwellian protons and massless cold electrons (with $\omega/kc \ll1$ assumption) is given by

\begin{equation}
\frac{\omega}{\Omega}-(k\lambda_i)^2+\frac{\omega}{\sqrt{2}|k|v_{\mathrm{th}}}Z(\zeta)+A\left(1+\zeta Z(\zeta)\right)=0.\label{kindisp}
\end{equation}
with $A=a-1,\,\zeta=\frac{\omega+\Omega}{\sqrt{2}|k|v_\mathrm{th}}$.
A solution with positive (negative) $\mathrm{Re}[\omega]$ solution corresponds to the right-handed (left-handed) polarized wave, respectively\citep{Stix1992}.
The dispersion relation for the CRC model has been derived in \citep{Jikei2021} where the Z-function is replaced by a rational-function approximation
\begin{equation}
\begin{split}
\frac{\omega}{\Omega}-(k\lambda_i)^2&-\frac{\omega}{\sqrt{2}|k|v_{\mathrm{th}}}\frac{(\pi-2)\zeta+i\sqrt{\pi}}{(\pi-2)\zeta^2+i\sqrt{\pi}-1}\\
&-\frac{A}{2}\frac{\pi-2}{(\pi-2)\zeta^2+i\sqrt{\pi}-1}=0. \label{CRCdisp}
\end{split}
\end{equation}
The dispersion relation of the CGL model can be obtained by linearizing the equations in subsection \ref{CGLeqs}.
The solution for parallel propagating transverse waves can be written in the following analytical form\citep{Hunana2019part1}
\begin{equation} \label{CGLdisp}
\begin{split}
\frac{\omega}{\Omega}=&\pm\frac{(k\lambda_i)^2}{2}\left[1+\beta_{\parallel}\left(1-\frac{a}{2}\right)\right]\\
&+k\lambda_i\sqrt{1+\frac{\beta_{\parallel}}{2}(a-1)+\frac{(k\lambda_i)^2}{4}\left[1-\beta_{\parallel}\left(1-\frac{a}{2}\right)\right]^2}.
\end{split}
\end{equation}
The condition $1+\frac{\beta_{\parallel}}{2}(a-1)<0$ is required for the CGL model to be unstable, which shows that this model cannot reproduce the EMIC instability ($\omega$ is always real if $a>1$).
The complex frequency in the CGL model always appears as a complex conjugate pair.
Therefore, it is clear that the CGL model lacks the cyclotron damping regardless of polarization.

\begin{figure}[H]
\begin{center}
\includegraphics[width=0.85\linewidth]{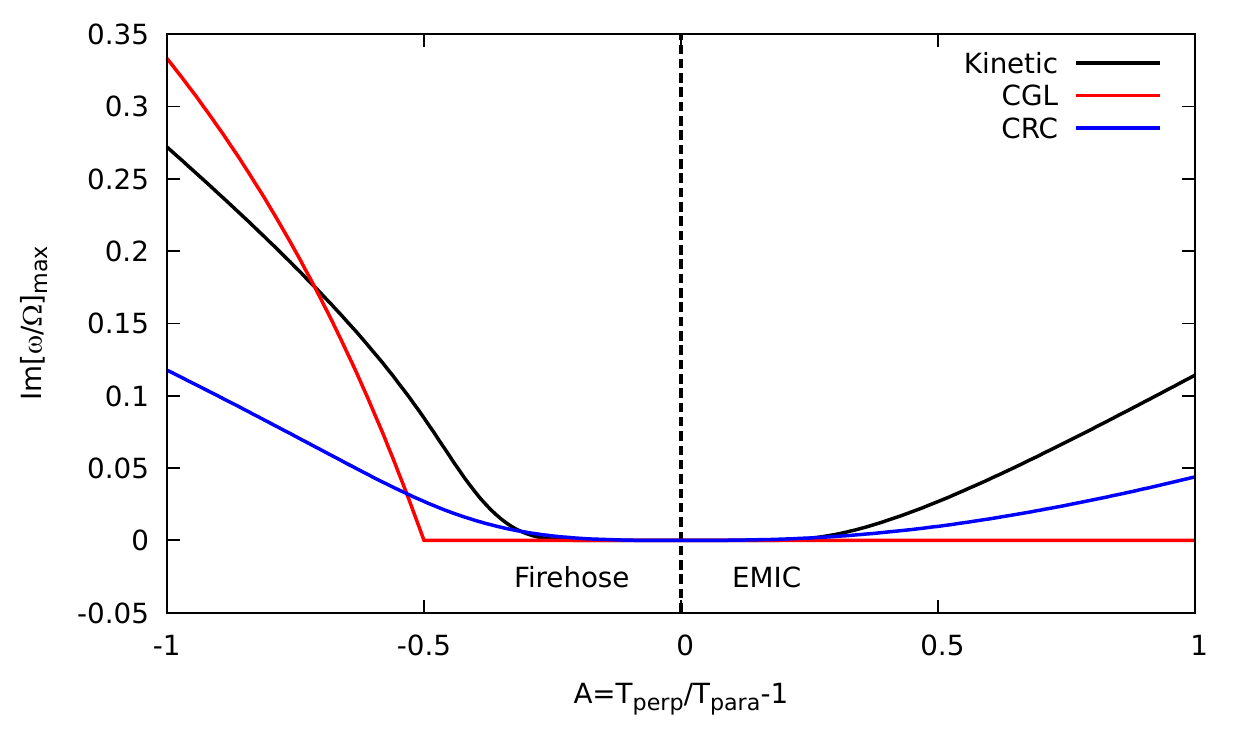}%
\caption{\label{mg} Maximum growth rate of each model as a function of $A=T_{\perp}/T_{\parallel}-1$ for a fixed parallel-beta fix of $\beta_{\parallel}=4$. The black line corresponds to the fully kinetic result, while the red and blue lines are obtained by the CGL and the CRC models, respectively. Note that the right-handed (left-handed) wave becomes unstable for $A<0$ ($A>0$).}
\end{center}
\end{figure}
FIG. \ref{mg} shows the maximum growth rate (the maximum value of $\mathrm{Re}[\omega(k)]$) of each model as a function of $A$ obtained with fixed parallel-beta $\beta_{\parallel}=4$.
While the CGL model may better reproduce the firehose instability, it completely misses the EMIC instability driven by cyclotron resonance.
On the other hand, the CRC model mimics the kinetic result for both the firehose and EMIC instability reasonably well.
Note that $A=-1$ corresponds to the $T_{\perp}=0$ limit, which gives the maximum growth rate for the firehose instability at a fixed $\beta_{\parallel}$. 
On the other hand, the growth rate for the EMIC instability increases beyond $A=1\,(T_{\perp}/T_{\parallel}=2)$.
For the purpose of comparison between the CRC model and the CGL model under unstable conditions, we will limit our discussion to the (parallel) firehose instability in the following.

\begin{figure}[H]
\begin{center}
\includegraphics[width=0.85\linewidth]{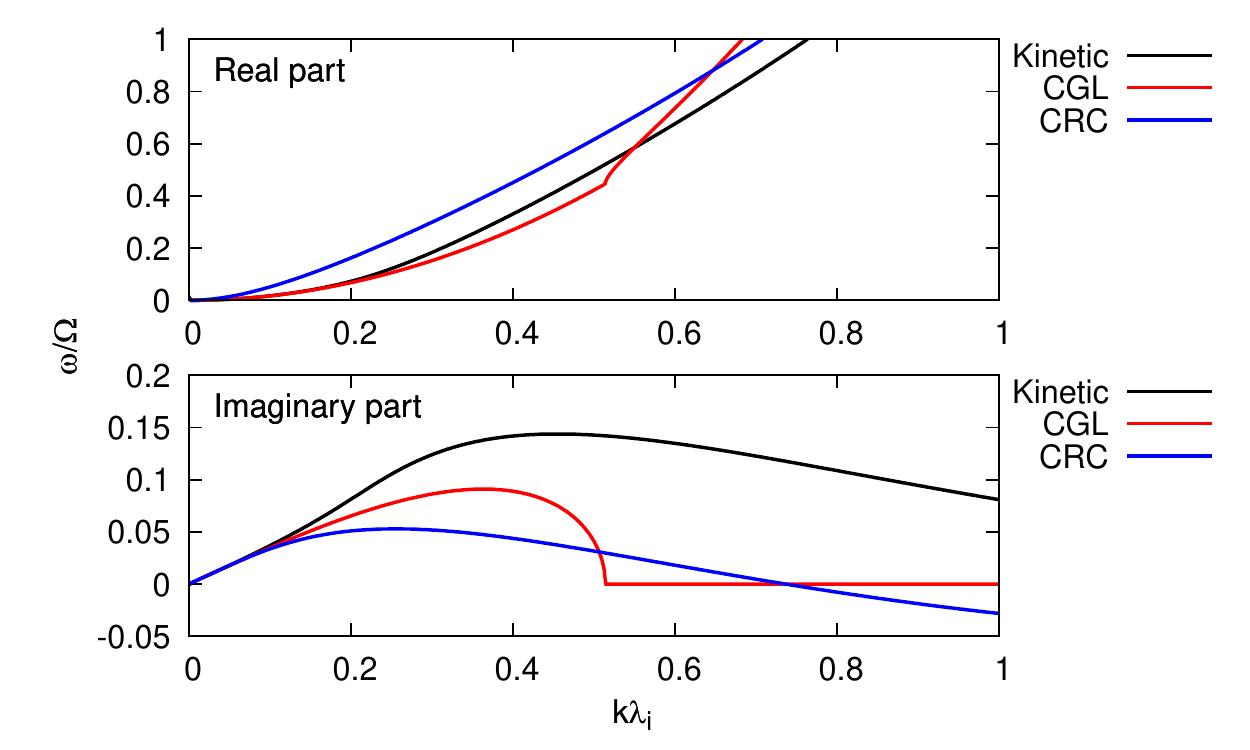}%
\caption{\label{disp2_5} Dispersion relation for $\beta_{\parallel}=2.5$. The top and bottom panels show the real and the imaginary parts of the frequency. The black, red, and blue lines indicate the fully kinetic result obtained by Eq. (\ref{kindisp}), the CGL model result by Eq. (\ref{CGLdisp}), and the CRC result by Eq. (\ref{CRCdisp}), respectively.}
\end{center}
\end{figure}

\begin{figure}[H]
\begin{center}
\includegraphics[width=0.85\linewidth]{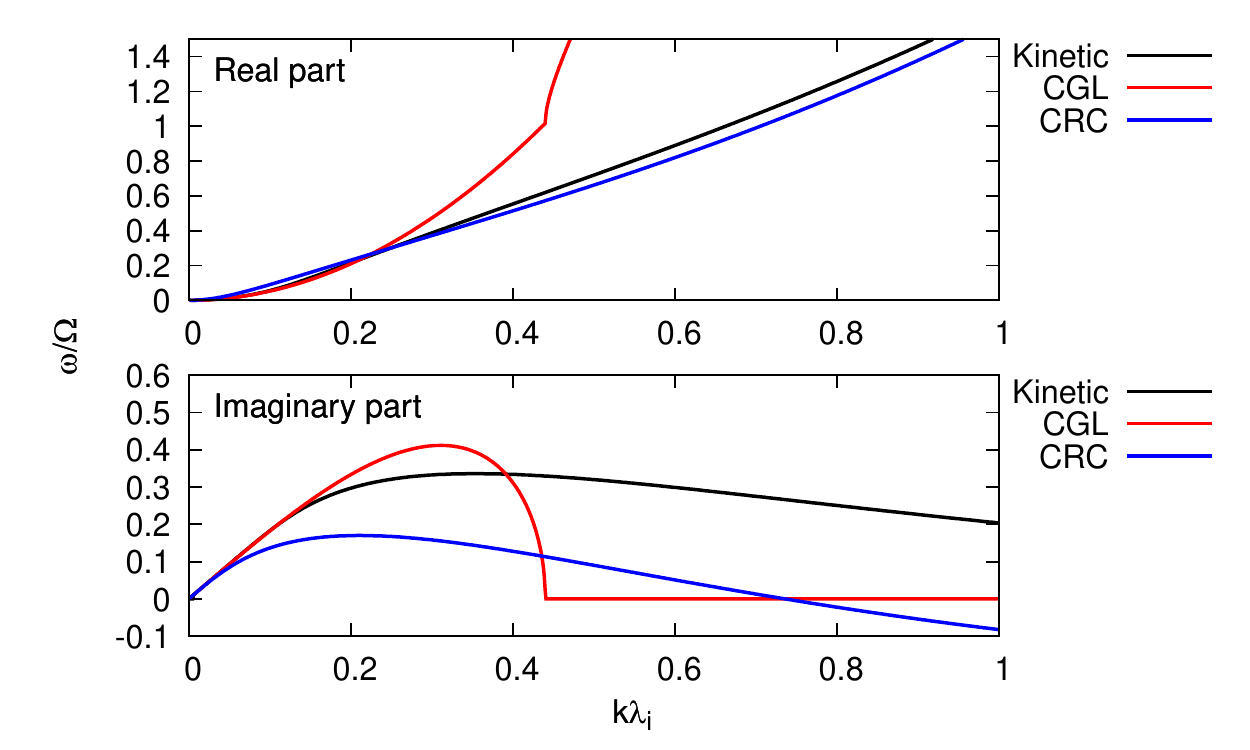}%
\caption{\label{disp10} Dispersion relation for $\beta_{\parallel}=10$ shown with the same format as FIG. \ref{disp2_5}}
\end{center}
\end{figure}

The dispersion relation of the right-handed wave predicted by each model is shown in Fig. \ref{disp2_5} and Fig. \ref{disp10}.
We show the results for two different parameters $\beta_{\parallel}=2.5,\,10$ with the same $a=0.1$, which we will use for nonlinear simulations in the next subsection.
Both models reproduce the instability in the long-wavelength limit.
Finite deviations are seen at the maximum growth rate and shorter wavelength.
In general, the maximum growth rate appears to be better described by the CGL model, which was also suggested by FIG. \ref{mg}.
However, it may be important to notice that the corresponding real frequency at around the maximum growth rate is not necessarily small, in apparent contradiction to the low-frequency approximation.
Also, the behavior of the CGL result changes in an indifferentiable way at the wavelength in which the frequency becomes purely real.
The CRC model typically gives a factor $\sim 2$ smaller maximum growth rate.
Instead, the smooth transition to a damping mode at a short wavelength qualitatively agrees with the fully kinetic result though the CRC damping rate is larger than the fully kinetic model.
We will see that the continuous transition between positive and negative growth rates in the CRC model plays a crucial role in the nonlinear evolution of the instability.

The error in the CGL model can be reduced by using better but more complicated FLR corrections\citep{Hunana2019part1}.
On the other hand, we expect that the error of the CRC model will be reduced if we extend the closure to higher-order moments.

\subsection{Simulation set up}
We compare the simulation results obtained with two different parameters for the parallel firehose instability discussed in \citet{Yoon2017}.
Both share the same temperaturature anisotropy ratio $T_{\perp}/T_{\parallel}=0.1$, whereas parallel plasma-betas are different between $\beta_{\parallel}=2.5$ and $\beta_{\parallel}=10$.
With a fully kinetic (particle-in-cell) simulation code and quasilinear theory, \citet{Yoon2017} found that the $\beta_{\parallel}=10$ case exhibits an almost quasilinear long term evolution.
On the other hand, the $\beta_{\parallel}=2.5$ case shows deviations from the quasilinear prediction based on the bi-Maxwellian distribution.
More specifically, the final state is quantitatively not consistent with the bi-Maxwellian prediction.
It was shown that wave-particle interaction results in a non-Maxwellian dumbbell-shaped distribution \citep{Seough2015,Yoon2017}.
Nevertheless, the nonlinear modification of the distribution function to a non-Maxwellian shape is clearly beyond the scope of our collisionless fluid models.

For all the simulations shown below, we use a simulation box size of $L_z=128\lambda_i$ with the grid size of $\Delta z=0.25\lambda_i$.
We use a time step of $\Omega\Delta t=0.01$ and 0.001 for the CRC and the CGL, respectively.
Although we have not done a precise numerical stability analysis, we empirically found that CGL model (especially with FLR correction) requires a smaller time step presumably due to the existence of fourth-order nonlinear terms $(P_{\parallel,\perp}\bm{\hat{b}}\cdot\nabla\bm{u}\cdot\bm{\hat{b}})$. 
It is also to be noted that the real frequency of short wavelength whistler wave is higher in the CGL than in the CRC model, which may also affect the numerical stability.
We use a Fourier pseudo-spectral method for spatial discretization and the fourth-order Runge-Kutta method for time evolution.
The initial condition is uniform in space with $10^{-4}B_0$ level of random noise added to the transverse magnetic field.

\subsection{Results}
\begin{figure}[H]
\begin{center}
\includegraphics[width=0.85\linewidth]{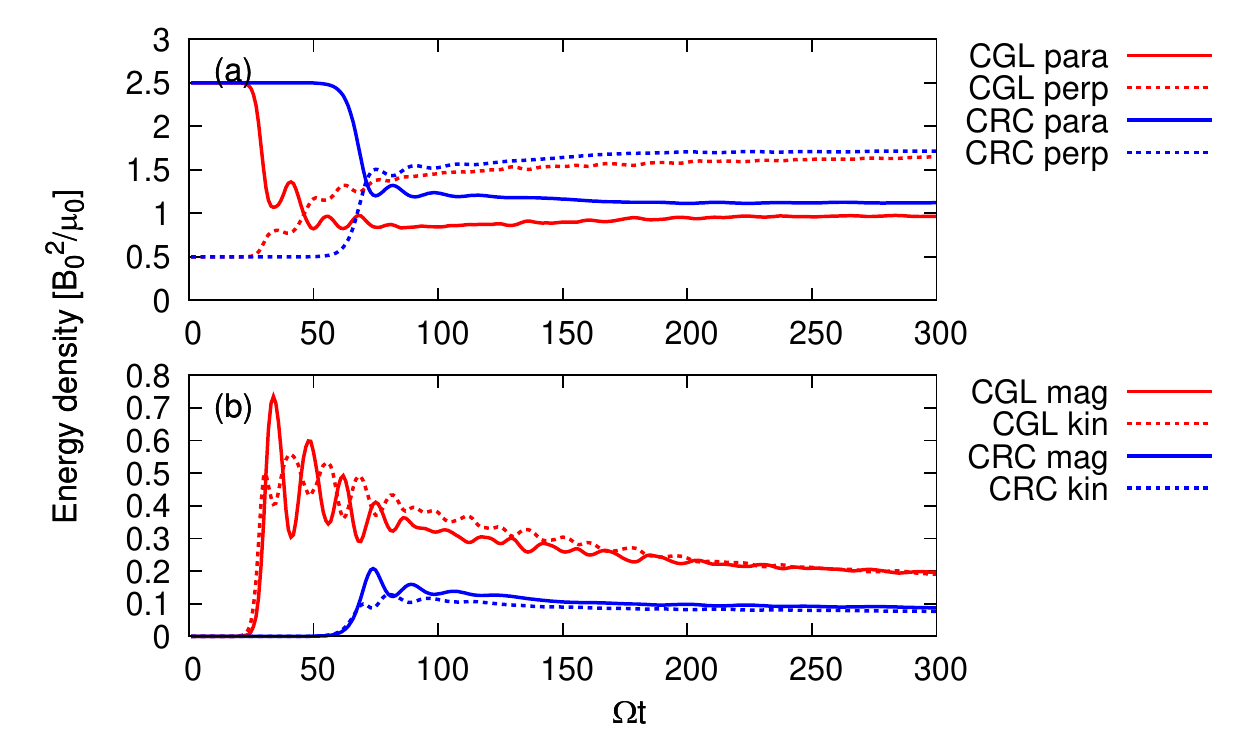}%
\caption{\label{energy10} Time evolution of energy for $\beta_{\parallel}=10$. In both panels, the red and blue lines show the CGL and the CRC model. The top panel (a) shows the spatial average of thermal energies. The solid and dotted lines corresponds to $P_{\parallel}/2$ and $P_{\perp}$ ($p_{zz}/2$ and $p_{xx}$). The bottom(b) shows the wave energy the solid lines are magnetic energy density $\frac{B_{x}^2+B_{y}^2}{2\mu_0}$ and the dotted lines are kinetic energy density $nm|\bm{u}|^2/2$. The sum of these four components are the value of Eq. (\ref{energy}) which conserves.}
\end{center}
\end{figure}

We first discuss the $\beta_{\parallel}=10$ results.
FIG. \ref{energy10} shows the time evolution of each energy component.
We identify the linear stage for the instability as $\Omega t\lesssim25$ for the CGL model and $\Omega t\lesssim60$ for the CRC model, respectively.
We have confirmed that magnetic energy growth shown in FIG. \ref{energy10} (b) is consistent with the maximum linear growth rates in FIG. \ref{disp10}.
After the linear stage, $25\lesssim\Omega t\lesssim30$ for the CGL model and $60\lesssim\Omega t\lesssim75$ for the CRC model, we can see that the parallel pressure is decreased and the perpendicular pressure is increased.
The isotropization during this phase is consistent with the quasilinear theory developed in subsection \ref{quasilinear}.
We thus call this stage the quasilinear phase.
After the quasilinear phase, $30\lesssim\Omega t$ for the CGL model and $75\lesssim\Omega t$ for the CRC model, the wave growth saturates, and the system evolves nonlinearly to an apparent quasi-steady state at around $\Omega t\sim200$ for the CGL model and $\Omega t\sim150$ for the CRC model.

\begin{figure}[H]
\begin{center}
\includegraphics[width=0.85\linewidth]{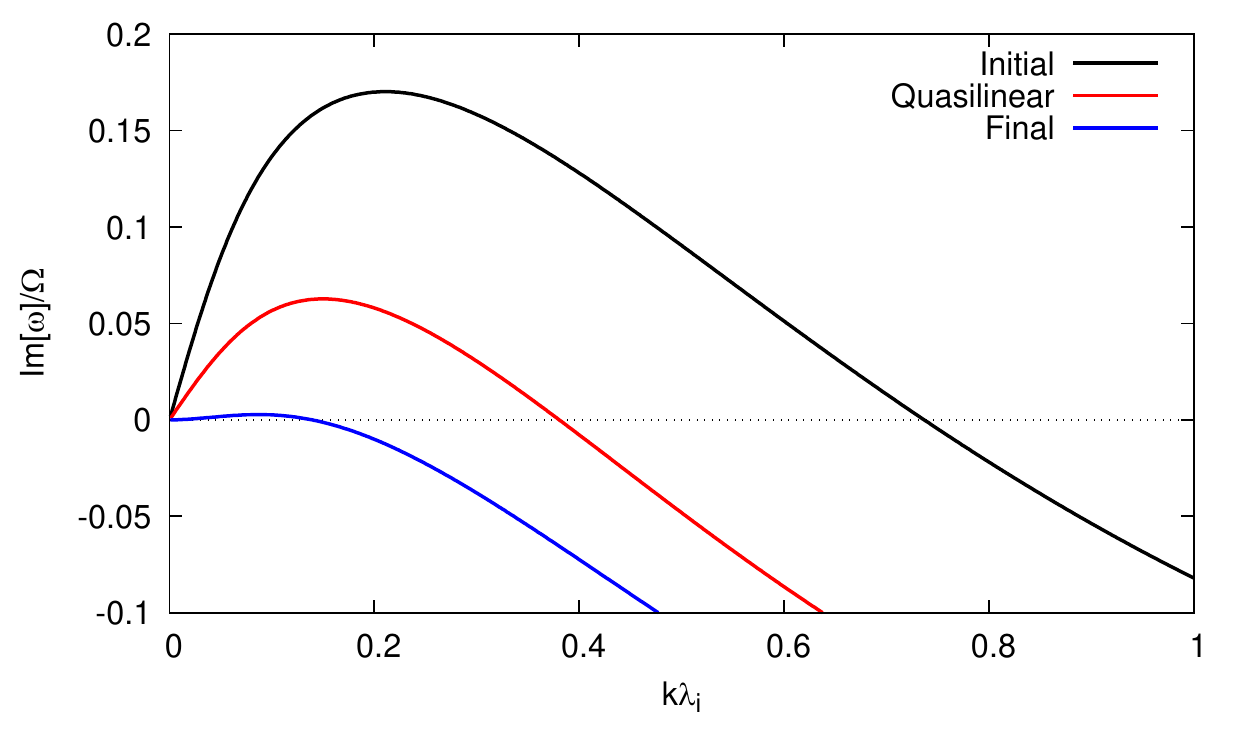}%
\caption{\label{ms} Evolution of linear growth rate for the CRC model. The black, red, and blue lines correspond to the initial condition $(\Omega t=0,\,\beta_{\parallel}=10,\,A=-0.9)$, late quasilinear phase $(\Omega t=66,\,\beta_{\parallel}=5.8,\,A=-0.55)$, and the final state $(\Omega t>200,\,\beta_{\parallel}=4.4,\,A=-0.22)$, respectively.}
\end{center}
\end{figure}

We find that the final state for the CRC model is approximately equal to the marginal stability in which the maximum linear growth rate becomes zero.
More specifically, the initial anisotropy $A=-0.9$ increases through the instability development to $A\sim-0.22$ at the final state.
The linear growth rates of the CRC model at different stages during the time evolution are shown in FIG. \ref{ms}.
As the system evolves through the quasilinear phase, the growth rate decreases associated with the decrease of $\beta_{\parallel}$ and the increase of $A$.
At the final state, we can see that the maximum growth rate is almost zero and that some initially unstable modes have shifted to damping modes.
On the contrary, we find that the final state of the CGL model is not consistent with the marginal stability.
The CGL model rather reduces the anisotropy more than that expected from the instability threshold condition $1+\frac{\beta_{\parallel}}{2}(a-1)=0$.
We attribute this to the lack of linear damping for the transverse waves in the CGL model.
Therefore, there is no reason why the linear growth rate for the transverse waves determines the final state.
Instead, as we see below, the damping for the transverse waves must involve the energy transfer to the sound mode wave that is subject to the Landau damping.
We will discuss the role of dissipation in more detail in the following. 

\begin{figure}[H]
\begin{center}
\includegraphics[width=0.85\linewidth]{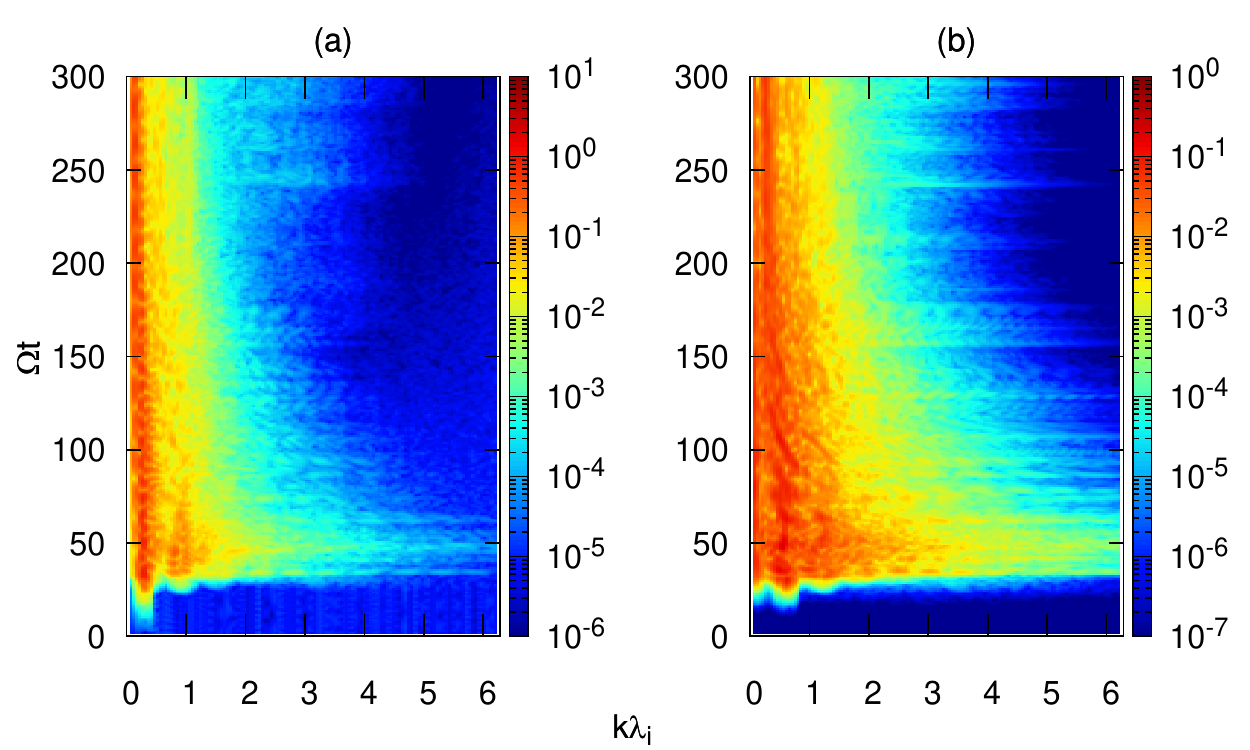}%
\caption{\label{CGLsp10} Time evolution of the wavenumber spectrum for the CGL model with $\beta_{\parallel}=10$. Pannel (a) is for the magnetic field $|\tilde{B}_x(k,t)|/B_0$ and (b) is for the density $|\tilde{n}(k,t)|/n_0$.}
\end{center}
\end{figure}

\begin{figure}[H]
\begin{center}
\includegraphics[width=0.85\linewidth]{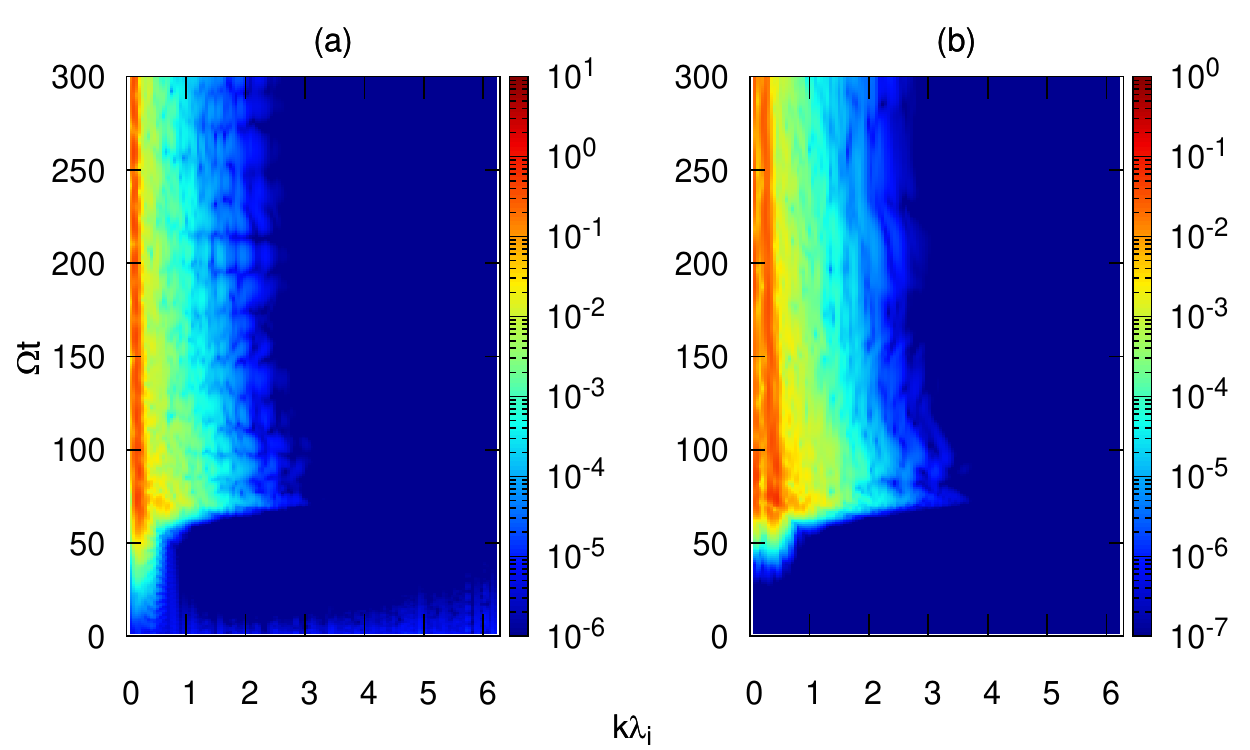}%
\caption{\label{NLsp10} Time evolution of the wavenumber spectrum for the CRC model with $\beta_{\parallel}=10$. The format is the same as FIG. \ref{CGLsp10}.}
\end{center}
\end{figure}

FIGs. \ref{CGLsp10}, \ref{NLsp10} show the time evolution of the wavenumber spectrum of each model.
Panel (a) shows the spectrum of magnetic field $|\tilde{B}_x(k,t)|$.
We can see that the unstable wavelength is consistent with FIG. \ref{disp10}.
In both cases, the power at higher wavenumbers appears in the quasilinear and early nonlinear phases, but they are eventually damped, and only long-wavelength modes are left in the final state.
It is noted that the power at short wavelengths in the late stage is larger in the CGL than the CRC, which may be understood by the damping rate shown in FIG. \ref{ms}.

Panel (b) shows the spectrum of the number density $|\tilde{n}(k,t)|$, which corresponds to the amplitude of sound waves.
The generation of density perturbation may be understood as a consequence of parametric decay instability (PDI)\citep{Goldstein1978,Terasawa1986}.
Through the PDI, a large amplitude transverse wave generates longitudinal oscillation via wave-wave interaction.
The parent electromagnetic wave distributes its energy to a co-propagating sound wave of a shorter wavelength and an electromagnetic wave of a longer wavelength.
In collisionless plasmas, the sound wave will eventually dissipate either via Landau damping or by steepening into a shock.
If the steepening dominates over the Landau damping, short-wavelength electromagnetic waves may also be generated nonlinearly by compression of transverse perturbations.
The CRC model has non-zero collisionless damping (even for right-handed waves), which leads to the dissipation of short-wavelength transverse fluctuations.
On the other hand, the dissipation in the CGL model always occurs through the Landau damping of longitudinal modes.
The high wavenumber modes in the CGL model should dissipate the energy indirectly through PDI and subsequent Landau damping.
While the indirect dissipation channel in the CGL model appears to be sufficient to balance with the energy injection in this specific parameter, this will not always be possible, as we see below.

\begin{figure}[H]
\begin{center}
\includegraphics[width=0.85\linewidth]{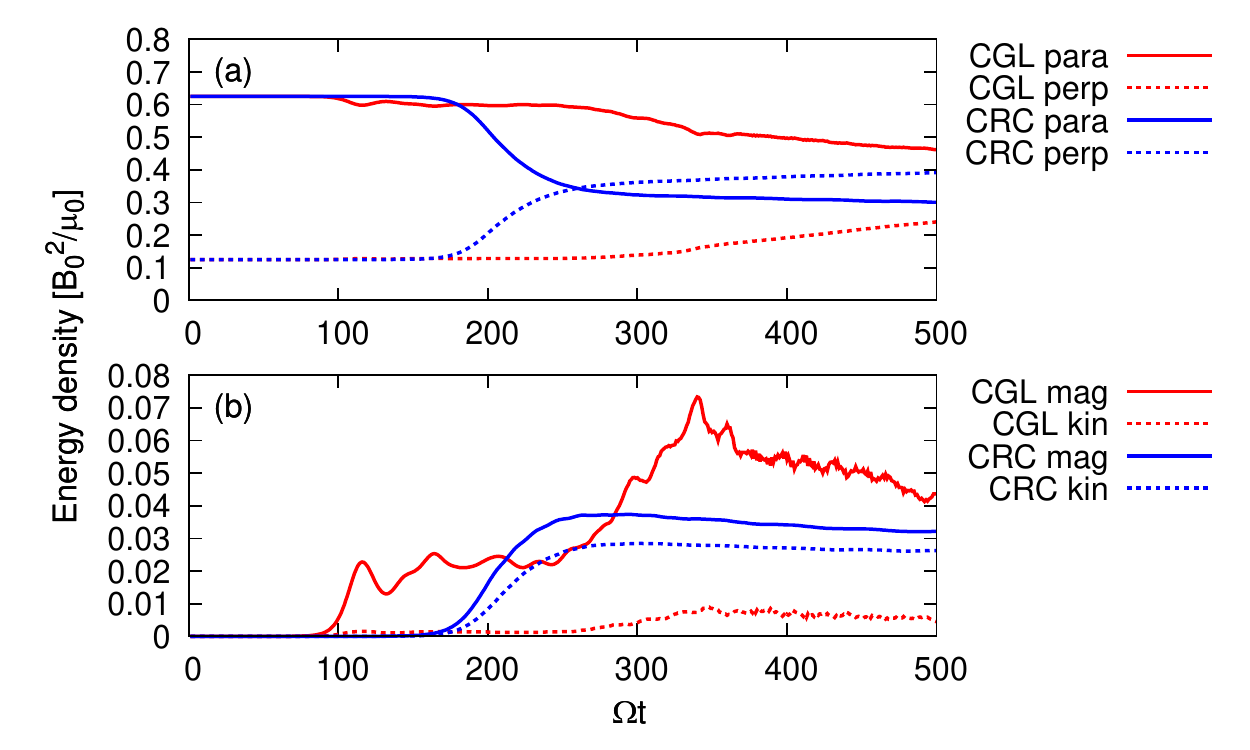}%
\caption{\label{energy2_5} Time evolution of energy for $\beta_{\parallel}=2.5$. The format is the same as in FIG. \ref{energy10}}
\end{center}
\end{figure}
Now we move on to the $\beta_{\parallel}=2.5$ case.
FIG. \ref{energy2_5} is the evolution of energy components.
For the CRC model, the qualitative behavior is the same as the $\beta_{\parallel}=10$ case.
The linear growth during $\Omega t\lesssim150$ (blue lines panel (b)) is consistent with FIG. \ref{disp2_5} . 
The system evolves quasilinearly during $150\lesssim\Omega t\lesssim250$ (blue lines in panel (a)), and the final state is consistent with the marginal stability analysis.

For the CGL model, the linear growth $(\Omega t\lesssim90)$ is as expected from the linear analysis, and quasilinear isotropization $(90\lesssim\Omega t\lesssim110)$ is also reasonable.
However, the nonlinear evolution is very different and appears to be physically unreasonable.
After exhibiting oscillatory behavior during $110\lesssim\Omega t\lesssim250$ the wave energy starts to increase again associated with the further reduction of anisotropy, which is not observed in the CRC model or fully kinetic simulations.
Note that we have confirmed that the aliasing effect at short wavelengths is not significant with the spatial resolution used in this study. We thus believe the behavior is not a numerical artifact.

\begin{figure}[H]
\begin{center}
\includegraphics[width=0.85\linewidth]{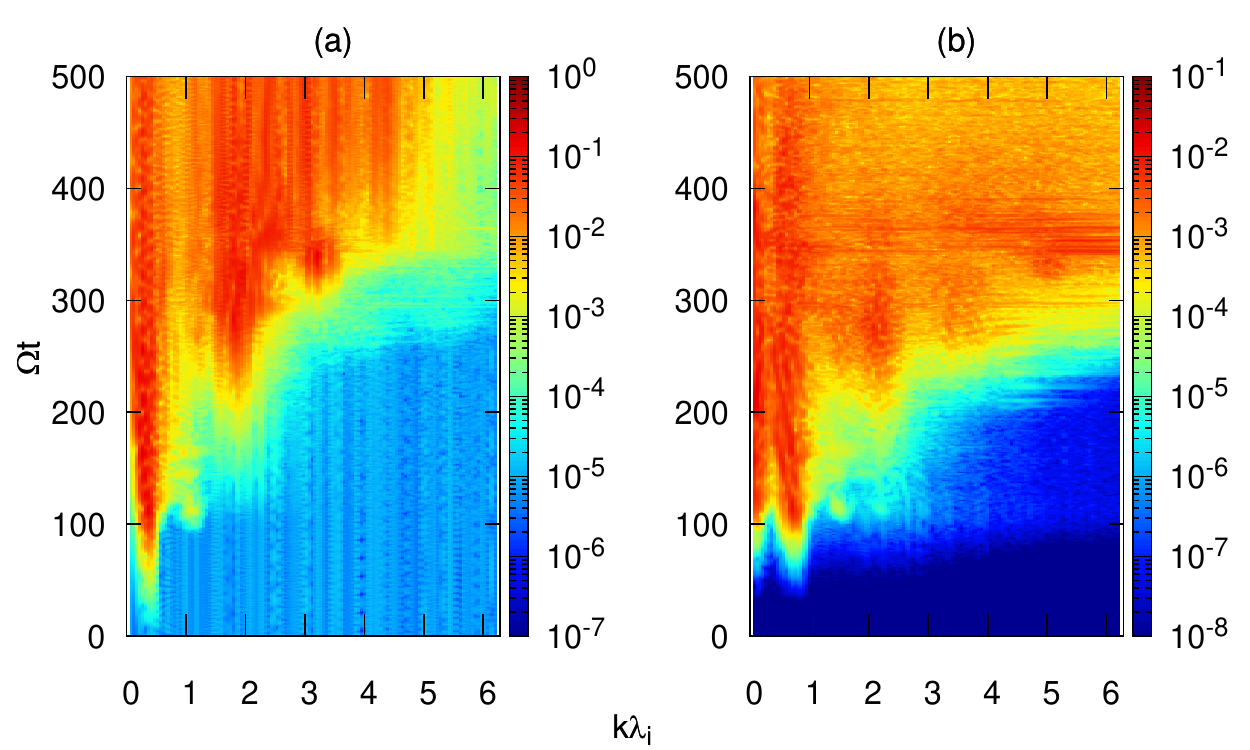}%
\caption{\label{CGLsp2_5} Time evolution of the wavenumber spectrum for the CGL model with $\beta_{\parallel}=2.5$. The format is the same as in FIG. \ref{CGLsp10}.}
\end{center}
\end{figure}

\begin{figure}[H]
\begin{center}
\includegraphics[width=0.85\linewidth]{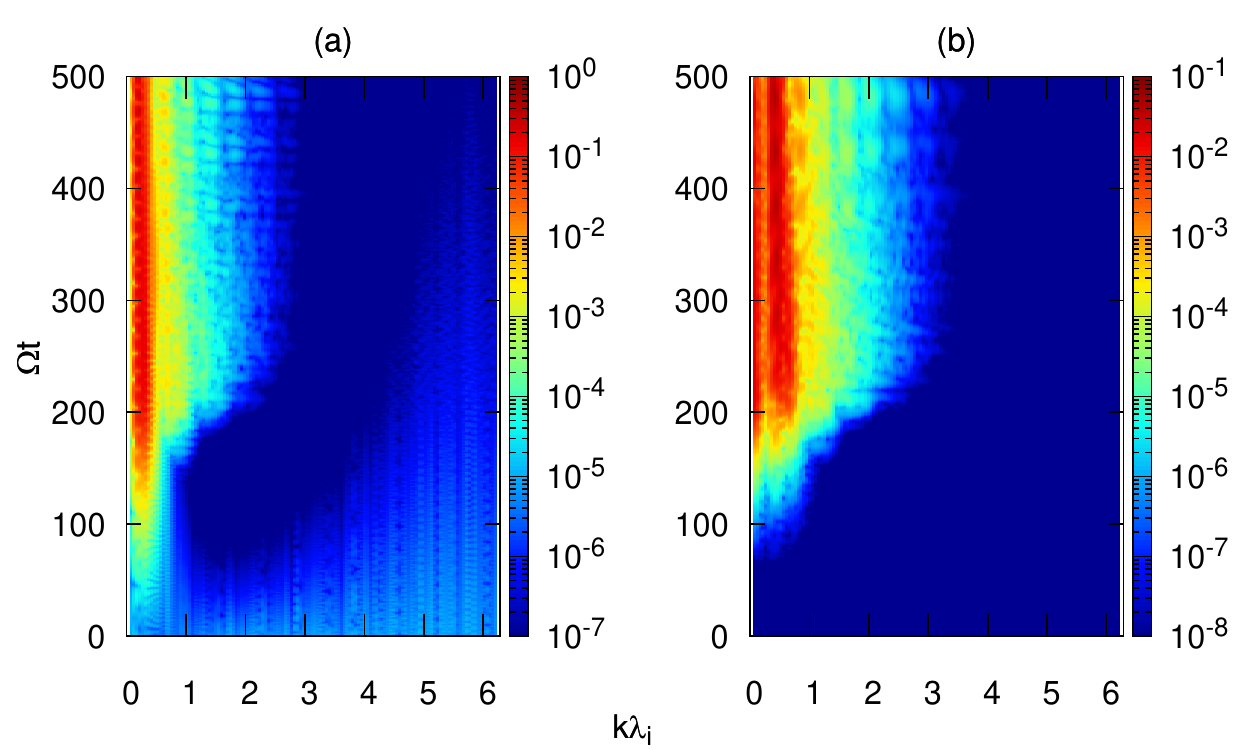}%
\caption{\label{NLsp2_5} Time evolution of the wavenumber spectrum for the CRC model with $\beta_{\parallel}=2.5$. The format is the same as in FIG. \ref{CGLsp10}.}
\end{center}
\end{figure}

The different behaviors are more clearly seen in the spectral evolution.
The CRC results (FIG. \ref{NLsp2_5}) are qualitatively the same as $\beta_{\parallel}=10$ case (FIG. \ref{NLsp10}).
On the other hand, the CGL result (FIG. \ref{CGLsp2_5}) indicates that undamped short wavelength may be the reason for the unphysical behavior.
At around  $\Omega t\sim250$, we can see the existence of large-amplitude, short-wavelength electromagnetic waves (FIG. \ref{CGLsp2_5} (a)).
We have confirmed the real frequency of whistler mode at $k\lambda_i\sim2$ with this parameter is about $9\Omega$ in the linear analysis, which clearly violates the low-frequency assumption made in the CGL model.
When a high-frequency wave exists in the CGL model, the pressure tensor is adjusted instantaneously to the FLR tensor $\Pi$ with respect to the oscillating magnetic field direction $\hat{\bm{b}}$.
This is physically unrealistic because the adjustment should occur over the time scale of the order of gyroperiod.
While this unrealistic assumption itself does not immediately imply a problem, it could introduce unphysical behaviors in the presence of large-amplitude fluctuations.

\begin{figure}[H]
\begin{center}
\includegraphics[width=0.85\linewidth]{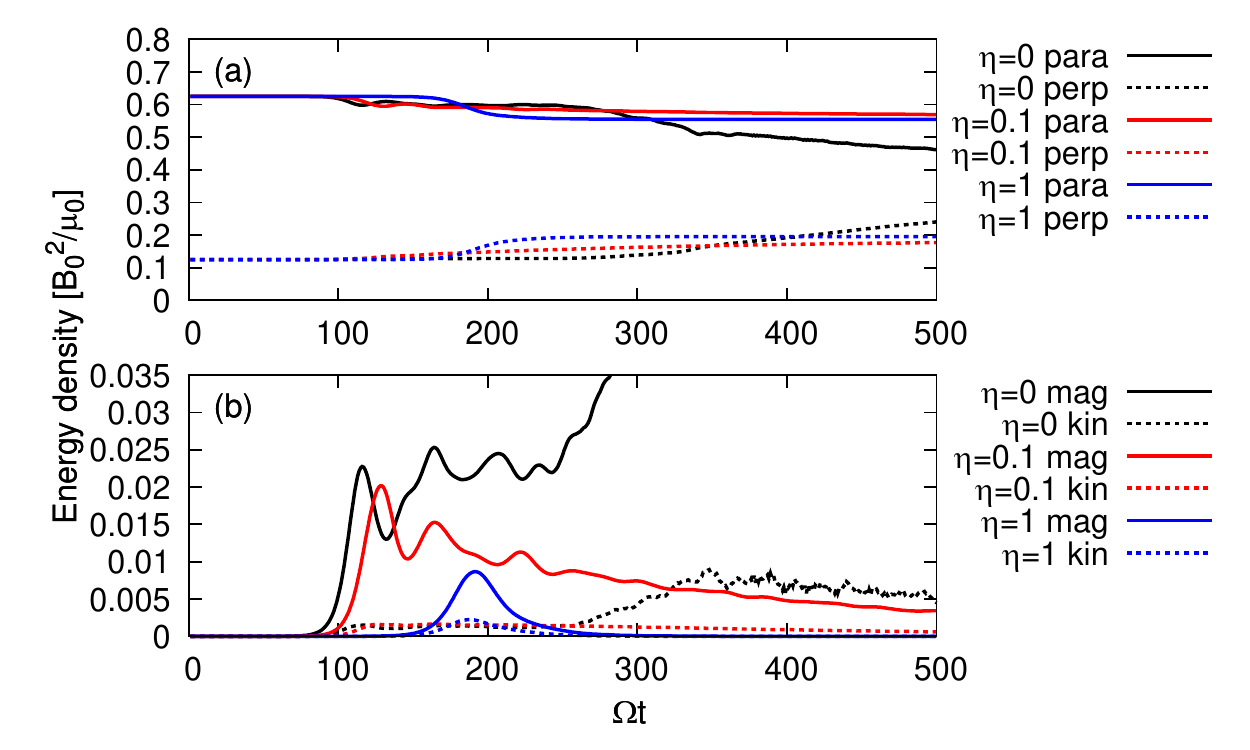}%
\caption{\label{energyres} Time evolution of energy for $\beta_{\parallel}=2.5$. The notations are the same as FIG. \ref{energy10} except for the colors. Black, red, and blue lines correspond to $\eta=0,\,0.1$ and 1 respectively.}
\end{center}
\end{figure}

To confirm that the short wavelength $(k\lambda_i\sim2)$ modes are responsible for the faulty behavior in the CGL model, we run the same simulation with the finite-resistivity CGL model.
FIG. \ref{energyres} shows the time evolution of energy for the CGL model with the same $\beta_{\parallel}=2.5,\,T_{\perp}/T_{\parallel}=0.1$ and three different restivities $\eta=0,\,0.1$ and 1.
Note that the resistive dissipation scale length is given by $l_d\sim\eta\lambda_i$.
This implies that the dissipation becomes significant at the grid scale $l_d\sim \Delta z$ for $\eta=0.1$, whereas a non-negligible impact apperars even for the linearly unstable mode for $\eta=1$ because of $l_d\sim\lambda_i$.
The $\eta=0.1$ case shows similar evolution as the non-resistive case during the linear and nonlinear phases, although the growth rate is slightly smaller.
However, the unphysical behavior in the nonlinear phase is not observed.
For the $\eta=1$ case, both the growth rate and the saturation level are much smaller, suggesting that this level of effective resistivity is too strong for modeling the collisionless plasma dynamics.

\begin{figure}[H]
\begin{center}
\includegraphics[width=0.85\linewidth]{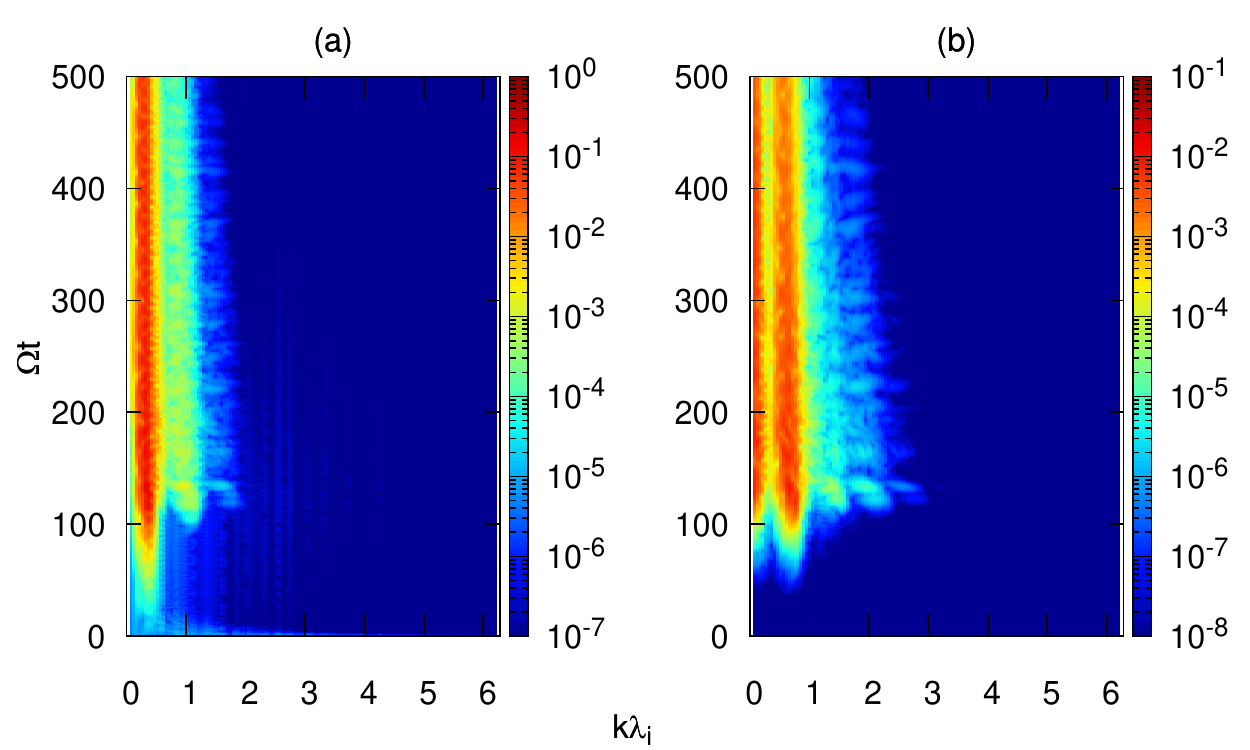}%
\caption{\label{spe0_1} Time evolution of the wavenumber spectrum for the CGL model with $\eta=0.1,\beta_{\parallel}=2.5$. The notations are the same as in FIG. \ref{CGLsp10}.}
\end{center}
\end{figure}

FIG. \ref{spe0_1} shows the wavenumber spectrum for the $\eta=0.1$ case.
We see that $k\lambda_i>2$ magnetic fluctuations are damped by the resistivity.
Consequently, the power at short wavelengths is significantly suppressed for both longitudinal and transverse modes.
We have confirmed that $\eta=0.1$ does not significantly affect the linear dispersion relation while suppressing the high-frequency waves.

Our simulation results indicate that the dissipation plays the central role in regulating the nonlinear development of the instability. Since the CRC model incorporates the collisionless damping both for the transverse and longitudinal modes, the final state as a result of the nonlinear evolution may be reasonably understood with the standard marginal stability. In contrast, the lack of physical dissipation of the transverse fluctuations in the CGL model may lead to substantially different consequences. In the absence of resistivity, the dissipation occurs indirectly through the energy transfer to longitudinal perturbations. Since the process is genuinely nonlinear, it would be difficult to understand the final state with a simple physical argument. Although a finite resistivity introduces linear damping, this makes the nonlinear evolution dependent on the assumed resistivity. Such a model is clearly not ideal for the purpose of modeling the collisionless system.

\section{Discussions and Conclusions} \label{sec4}
In this paper, we have compared two collisionless fluid models that incorporate linear kinetic effects.
The CGL model based on the pressure tensor decomposition correctly predicts the dispersion relation of long-wavelength, low-frequency waves in collisionless plasmas.
The CRC model calculates the time evolution of the full pressure tensor and explicitly takes into account the non-local heat flux tensor associated with transverse perturbations.
The CGL model is better suited for low-frequency (relative to the proton cyclotron frequency) phenomena such as the parallel firehose instability in the sense it reproduces the fully kinetic dispersion relation more accurately than the CRC model.
On the other hand, the CRC model incorporates the cyclotron resonance effect and can be used for higher frequency phenomena, including the EMIC instability.
While it is less accurate than the CGL model for the low-frequency regime, it provides an adequate approximation for a much wider range of parameters, in particular where the cyclotron resonance plays an essential role.

We performed fully nonlinear simulations of initially firehose-unstable systems with two different parameters to investigate whether these models are capable of simulating the nonlinear evolution of collisionless plasmas.
The CRC model reproduces much of the proton kinetic effect except for the deformation of the distribution function to a non-bi-Maxwellian shape.
The linear and quasilinear stages are consistent with the theoretical predictions, and the system evolves nonlinearly into the final quasi-steady states, which can be described by the marginal stability.

The CGL model, on the other hand, cannot directly dissipate transverse waves by the cyclotron damping.
Therefore, the final state of the CGL model is, in general, not consistent with the marginal stability condition.
If the Landau damping is sufficiently strong to dissipate the sound waves generated by PDI and subsequent steepening ($\beta_{\parallel} = 10$), the balance between the generation and damping of short-wavelength fluctuations determines the final state. In the opposite case, steepening of the short-wavelength compressional perturbations may lead to a more complicated nonlinear evolution. For the case $\beta_{\parallel} = 2.5$, short-wavelength and high-frequency magnetic fluctuations are generated as a result. The dissipation of these waves appears to be insufficient, which results in the unphysical behavior. We suggest that it may be, in part, due to the high-frequency nature of the fluctuations, which violates the low-frequency approximation of the model. It should be noted that the dissipation of transverse waves in the CGL model occurs only through the nonlinear energy transfer to longitudinal modes (via PDI), which is fundamentally different from the linear cyclotron damping.

Although it is possible to filter out those short-wavelength magnetic field fluctuations by adding an artificial resistivity, this approach is not ideal for the purpose of modeling collisionless plasmas for the following reasons.
First, it is not possible to accurately determine the effective resistivity in collisionless plasmas without taking into account resonant wave-particle interaction.
The dissipation of transverse waves depends not only on the field quantities such as the plasma beta and anisotropy but also on the polarization and wavelength.
Therefore, the resistivity cannot be written in a simple form.
Second, the way in which the effective Ohmic heating distributes energy to the pressure tensor components cannot be determined self-consistently.
In reality, the amount of energy that the parallel and the perpendicular components absorb from the wave is not the same if the cyclotron resonance is the origin of the dissipation.
Furthermore, electrons may also be responsible for dissipation, in particular, at short wavelengths.
In this case, the electrons should also absorb the dissipated energy in a non-trivial way.

While collisionless fluid models have often been evaluated in terms of the reproducibility of kinetic instability, our results suggest that the collisionless damping, both for the longitudinal and transverse modes, plays an important role in the nonlinear evolution. Based on the finding, we conclude that the CRC model that naturally incorporates the collisionless damping will be better suited for a nonlinear simulation model.

Now let us discuss modifications by the electrons, which were treated as cold and massless in most of this paper.
In principle, we could treat the electrons as a separate fluid component in both the CGL and the CRC models.
However, we saw that the CGL model could fail at a scale much longer than the electron inertial length.
Therefore, extending the CGL model to smaller scales will not be straightforward.
For the CRC model, we can use the same approach for the protons and electrons independently.
We can thus take into account resonant wave-particle interaction for the two species at the same time.
For instance, such a model will be able to reproduce the whistler instability driven by an anisotropic electron distribution $T_{\perp e}/T_{\parallel e}>1$.
Also, the longitudinal waves now have a characteristic length scale, which is determined by the Debye length.
The Langmuir waves and the (dispersive) sound waves will be strongly damped at this scale. 

Finally, we will mention three remaining issues in the CRC model.
First of all, the linear dispersion relation of the CRC model sometimes contains substantial deviations from the fully kinetic model, which may not be acceptable in practical applications.
Specifically, the maximum growth rate of temperature anisotropy instability is smaller compared to the fully kinetic result, typically by a factor of 2 to 3.
We can improve the accuracy by continuing the moment expansion of the Vlasov equation to higher orders to construct a more accurate non-local closure.
This corresponds to giving more degrees of freedom to the rational-function approximation of the plasma response.
Second, the CRC at this point is valid only for parallel propagation.
We could use the same kind of closure to oblique propagation with some corrections to include the effect of finite $k_{\perp}$ for better accuracy ($k_{\perp}$ is the wavenumber perpendicular to the ambient magnetic field).
Third, the CRC model currently uses Fourier transform, which can only be performed on periodic systems.
For non-periodic boundary conditions or in very large systems, a finite interval approximation of the Hilbert transform may be needed.
However, this may be more of a numerical issue rather than a theoretical problem.
It is clear that we still need an effort to resolve these issues.
Nevertheless, we believe that the CRC model is a promising candidate that can be used in a wide range of nonlinear simulations to achieve better accuracy than conventional fluid models while demanding much less computational resources than fully kinetic simulation models.

\begin{acknowledgments}
This work was supported by JSPS KAKENHI Grant Nos.~17H02966 and 17H06140. T.~J. was supported by the International Graduate Program for Excellence in Earth-Space Science (IGPEES), The University of Tokyo.
\end{acknowledgments}

\section*{DATA AVAILABILITY}
The data that support the findings of this study are available from the corresponding author upon reasonable request.

\appendix

\nocite{*}
\providecommand{\noopsort}[1]{}\providecommand{\singleletter}[1]{#1}%

\end{document}